\documentclass[a4paper,11pt]{article}
\usepackage{amssymb}
\usepackage{amsmath}
\usepackage{xspace}
\usepackage{caption}
\usepackage{appendix}
\usepackage{float}
\usepackage{hyperref}
\usepackage{diagbox}
\usepackage{soul}
\usepackage{subfig}
\usepackage{slashed}
\usepackage{amsfonts}
\usepackage{multirow}
\usepackage{mathrsfs}
\usepackage{graphicx}
\usepackage{amsmath}
\usepackage{amssymb}
\usepackage{bm}
\usepackage{bbm}
\usepackage{color}
\usepackage{tcolorbox,colortbl}
\usepackage{booktabs}
\usepackage{comment}
\usepackage{slashed}
\usepackage{geometry}
\usepackage{authblk}
\usepackage[numbers,sort&compress]{natbib}

\newcommand{\M}{\ensuremath \text{M}}
\newcommand{\als}{\alpha_s}
\newcommand{\eps}{\epsilon}

\allowdisplaybreaks[4]
\title{Higgs boson decay to massive bottom quarks at order $\alpha_s^4$\\
 induced by top-quark Yukawa couplings}

\author[3,4]{Jian Wang}
\author[5,6]{Xing Wang}
\author[1,2]{Yefan Wang}
\affil[1]{Department of Physics and Institute of Theoretical Physics, Nanjing Normal University, Nanjing, Jiangsu 210023, China}
\affil[2]{Nanjing Key Laboratory of Particle Physics and Astrophysics, Nanjing Normal University, Nanjing,
Jiangsu 210023, China}
\affil[3]{School of Physics, Shandong University, Jinan, Shandong 250100, China}
\affil[4]{Center for High Energy Physics, Peking University, Beijing 100871, China}
\affil[5]{School of Science and Engineering, The Chinese University of Hong Kong, Shenzhen, Longgang, Shenzhen, Guangdong 518172, China}
\affil[6]{Southern Center for Nuclear-Science Theory (SCNT), Institute of Modern Physics,
Chinese Academy of Sciences, Huizhou, Guangdong 516000, China
}

\date{March 23, 2026}

\begin{document}

\maketitle

\begin{abstract}
The Higgs boson decay to massive bottom quarks has the largest branching ratio.
The decay is mainly induced by the bottom-quark Yukawa coupling with the decay rate calculated up to $\mathcal{O}(\alpha_s^4)$ assuming the massless final-state bottom quark.
The top-quark Yukawa coupling induced contribution starts at $\mathcal{O}(\alpha_s^2)$, and exhibits logarithmic and power enhancements, making the perturbative expansion converge slowly,
which is a feature not present in the hadronic Higgs boson decay.
We present a calculation of such contributions at $\mathcal{O}(\alpha_s^4)$ to the decay into massive bottom quarks in which the squared amplitudes contain two top-quark Yukawa couplings and the final state must include at least a bottom quark pair.
We find that they increase the decay width, relative to the result up to $\mathcal{O}(\alpha_s^3)$, by  $0.4\%$, larger than the experimental precision at future lepton colliders, and reduce the scale dependence significantly down to $0.4\%$.

\end{abstract}

\newpage
\section{Introduction}
The decay of the Higgs boson into a bottom--antibottom quark pair, $H \to b\bar{b}$, is the dominant channel in the Standard Model (SM), 
governing the Higgs boson total width and providing a direct probe of the bottom-quark Yukawa coupling. 
Precision measurements of this coupling, projected to reach the percent or even subpercent level at future colliders such as the HL-LHC  \cite{ATLAS:2018jlh} and a potential $e^+e^-$ Higgs factory \cite{Zhu:2022lzv,CEPCPhysicsStudyGroup:2022uwl,Altmann:2025feg}, necessitate a theoretical prediction with matching precision. 
The primary source of theoretical uncertainty stems from higher-order Quantum Chromodynamics (QCD) corrections, which are substantial due to the large strong coupling constant at the relevant scale and the emergence of high-energy logarithms.

A systematic perturbative expansion is essential for a reliable prediction. 
The calculation of this decay width has a long history, with the next-to-leading order (NLO) QCD and electroweak (EW)  corrections established thirty years ago \cite{Braaten:1980yq,Sakai:1980fa,Janot:1989jf,Drees:1990dq,Dabelstein:1991ky,Kniehl:1991ze}.
The next-to-next-to-leading order (NNLO) QCD corrections were calculated for the total and differential decay rate in \cite{Chetyrkin:1995pd,Harlander:1997xa,Primo:2018zby,Wang:2023xud} and \cite{Bernreuther:2018ynm,Behring:2019oci,Somogyi:2020mmk}, respectively. 
Even higher order QCD corrections were obtained assuming that the final-state bottom quark is massless \cite{Gorishnii:1990zu,Chetyrkin:1996sr,Baikov:2005rw,Herzog:2017dtz,Anastasiou:2011qx,DelDuca:2015zqa,Mondini:2019gid,Chen:2023fba,Yan:2024oyb,Fox:2025cuz}. 
Simultaneously, the mixed QCD$\times$EW corrections were computed in  \cite{Kataev:1997cq,Mihaila:2015lwa}.
Recently, the top quark induced contribution has been investigated at $\mathcal{O}(\alpha_s^2)$ \cite{Primo:2018zby} and $\mathcal{O}(\alpha_s^3)$ \cite{Mondini:2020uyy,Wang:2024ilc}, respectively\footnote{The top quark induced contribution to the hadronic Higgs boson decay, 
which can include $b\bar{b}$ or $gg$ in the final state, has been calculated up to $\mathcal{O}(\alpha_s^4)$  \cite{Chetyrkin:1997vj,Davies:2017xsp}.
}. 
The calculation in \cite{Wang:2024ilc} was performed within an effective field theory framework, integrating out the top quark, and decomposing the width into contributions from different effective operator combinations: $C_2C_2$ (pure bottom Yukawa), $C_1C_2$ (interference), and $C_1C_1$ (pure gluonic),
where $C_i$ are the Wilson coefficients of the effective operators.
For the dominant $C_2C_2$ contribution, corrections beyond NNLO were found to be below $0.2\%$ when assuming massless final-state $b$-quarks \cite{Baikov:2005rw,Herzog:2017dtz}. 
However, the $C_1C_2$ interference term, induced by the top-quark Yukawa coupling, exhibits a markedly different behaviour. 
It is enhanced by large logarithmic terms $\log^j(m_H^2/m_b^2)$ with $j=1,2$ already at $\mathcal{O}(\alpha_s^2)$, and at $\mathcal{O}(\alpha_s^3)$ it includes logarithmic terms up to $\log^4(m_H^2/m_b^2)$. 
These logarithms, originating from soft massive quark effects at subleading power, are not Sudakov-like and are characterised by a distinct colour structure proportional to $C_A - C_F$. 
Note that there are no such logarithms in the decay to all hadronic states, where the limit $m_b\to 0$ works well \cite{Chetyrkin:1997vj,Davies:2017xsp}.
Although the $C_1C_1$ contribution appears first at $\mathcal{O}(\alpha_s^3)$,
it is power enhanced with respect to the $C_2C_2$ and $C_1C_2$ terms.
Consequently, the $\text{N}^3\text{LO}$ corrections increase the NNLO result by approximately $1\%$, significantly exceeding naive $\alpha_s$ power-counting expectations \cite{Wang:2024ilc}. 
This underscores the importance of a full analytic treatment of finite $b$-quark mass effects in the top-induced contribution for precise phenomenology.

With the $\text{N}^3\text{LO}$ calculation complete, the logical next step is the computation of the fourth-order ($\text{N}^4\text{LO}$) corrections. This advancement is crucial for several reasons. First, it will further reduce the residual renormalization scale dependence, which remains one of the dominant theoretical uncertainties at $\text{N}^3\text{LO}$. Second, it provides a decisive test of the convergence of the perturbative series at an unprecedented level. Third, it probes the higher-order structure of the soft massive quark logarithms that appear in the $C_1C_2$ and $C_1C_1$ channels, offering essential data for developing all-order resummation formalisms at subleading power.

Since the bottom quark mass effect is negligible in the $C_2C_2$ channel, 
it is reasonable to take the result in refs. \cite{Baikov:2005rw,Herzog:2017dtz} with massless bottom quarks.
In this paper, as an extension of our previous works \cite{Wang:2023xud,Wang:2024ilc,Wang:2025son},  we present an analytic result of the $\mathcal{O}(\alpha_s^4)$ correction in the $C_1C_1$ channel with full dependence on the bottom quark mass, leaving the calculation of the $C_1C_2$ channel to future work.

This paper is organized as follows.
In section \ref{sec:frame}, we describe the framework to perform the calculation,
including the effective operators and the decomposition of the decay width into different channels.
The MIs needed in the amplitude of $H\to b\bar{b}\to H$ are calculated analytically in section \ref{sec:MIs},
while the numerical results of the decay width are provided in section \ref{sec:num}.
We conclude in section \ref{sec:con}.

\section{Calculation framework}
\label{sec:frame}

In calculation of the decay rate of $H\to b\bar{b}$, we adopt the effective Lagrangian
\begin{align}
\mathcal{L}_{\text{eff}} = -\frac{H}{v}\left(
C_1 \mathcal{O}_1^R
+C_2 \mathcal{O}_2^R
\right)+ \mathcal{L}_{\text{QCD}}\,,
\label{Leff}
\end{align}
where $v$ is the vacuum expectation value of the Higgs field $H$ and $\mathcal{L}_{\text{QCD}}$ is the QCD Lagrangian with the top quark decoupled.   
The two renormalized effective operators are defined by
\begin{align}
\mathcal{O}_1^R =  Z_{11}\mathcal{O}_1 + Z_{12} \mathcal{O}_2,\quad
\mathcal{O}_2^R =  Z_{21} \mathcal{O}_1 + Z_{22} \mathcal{O}_2
\label{mix}
\end{align}
with 
\begin{align}
\mathcal{O}_1 = 
(G^{0}_{a,\mu\nu})^2,\quad
\mathcal{O}_2= m_b^{0}\bar{b}^{0}{b}^{0}.
\end{align}
Here the superscript ``$0$'' indicates that the fields and couplings are bare quantities.
The mixing between these two effective operators is described by the renormalization constants $Z_{ij}$,
 which are given by \cite{Kluberg-Stern:1974iel,Nielsen:1975ph,Nielsen:1977sy,Chetyrkin:1996ke,Chetyrkin:1997un},
\begin{align}
Z_{11}&= 1+ \als\frac{\partial \log Z_{\als}}{\partial \als}  
=  1 + \left(\frac{\als}{\pi}\right)\left(\frac{-11C_A+2n_f}{12\eps}\right)
\nonumber\\&\quad
+\left(\frac{\als}{\pi}\right)^2\left(
\frac{\left(-11C_A+2n_f\right)^2}{144\eps^2}-\frac{17C_A^2-5C_An_f-3C_Fn_f}{24\eps}
\right)+\mathcal{O}(\als^{3}),\nonumber\\
Z_{12} &= -4\als\frac{\partial \log Z^{\overline{\text{MS}}}_m}{\partial \als}= \left(\frac{\als}{\pi}\right)\left( \frac{3C_F}{\eps}\right)
\nonumber\\&\quad
+\left(\frac{\als}{\pi}\right)^2C_F\left(
\frac{-11C_A+2n_f}{4\eps^2}+\frac{97C_A+9C_F-10n_f}{24\eps}
\right)+\mathcal{O}(\als^{3}),\nonumber\\
Z_{21} & =0, \nonumber\\
Z_{22} & =1,
\label{eq:renZ}
\end{align}
where $C_A=3$ and $C_F=4/3$ in QCD and $n_f=5$. 
The top quark effect is encoded in the Wilson coefficients  \cite{Chetyrkin:1996ke,Chetyrkin:1997un,Schroder:2005hy,Chetyrkin:2005ia,Liu:2015fxa,Davies:2017xsp},
\begin{align}
C_1 &= -\left(\frac{\als}{\pi}\right)\frac{1}{12}- \left(\frac{\als}{\pi}\right)^2\frac{11}{48} 
- \left(\frac{\als}{\pi}\right)^3\left[\frac{137}{576}L_t+\frac{443}{864}\right]+\mathcal{O}(\als^4),\\
C_2 &= 1 + \left(\frac{\als}{\pi}\right)^2\left[
- \frac{1}{3}L_t+\frac{5}{18}\right] 
+ \left(\frac{\als}{\pi}\right)^3 
\left[
-\frac{23}{36}L_t^2-\frac{79}{36}L_t+\frac{749}{1296}+\frac{5}{3}\zeta(3)
\right]
\nonumber\\&
\quad+\left(\frac{\als}{\pi}\right)^4
\bigg[
-\frac{529 }{432}L_t^3-\frac{1093}{144} L_t^2+\frac{55\zeta (3)}{4}L_t-\frac{14045}{864}L_t-\frac{575 \zeta (5)}{36}
+\frac{14}{3} \text{Li}_4\left(\frac{1}{2}\right)
\nonumber\\&\quad
-\frac{277 \pi ^4}{1440}-\frac{7\pi^2\log^2(2)}{36}+\frac{7 \log ^4(2)}{36}+\frac{30773 \zeta (3)}{1536}
-\frac{2\pi ^2 \log (2)}{27}-\frac{\pi ^2}{27}+\frac{198665}{62208}
\bigg]
+\mathcal{O}(\als^5),\nonumber
\label{C1C2}
\end{align}
where $L_t$ = $\log(\mu^2/m_t^2)$ and $m_t$ is the on-shell top quark mass.  
The $1/m_t^2$ higher power corrections in the decay width of $H\to b\bar{b}$ start at $\mathcal{O}(\alpha_s^2)$, modifying the $\mathcal{O}(\alpha_s^2)$ correction by only $0.1\%$ \cite{Primo:2018zby}, and thus can be fully neglected in the decay width.

The decay amplitude of $H\rightarrow b\bar{b}$ can be induced by either of the effective operators and thus the decay width can be decomposed as 
\begin{align}
\Gamma_{H\rightarrow b\bar{b}}=  \Gamma^{C_2C_2}_{H\rightarrow b\bar{b}} + \Gamma^{C_1C_2}_{H\rightarrow b\bar{b}} + \Gamma^{C_1C_1}_{H\rightarrow b\bar{b}}.
\end{align}
The superscripts denote the combination structure of effective operators in the squared amplitudes.
Each term in the above equation can be expanded in the strong coupling $\als$, 
\begin{align}
\Gamma^{C_2C_2}_{H\rightarrow b\bar{b}} &= C_2C_2\left[\Delta^{C_2C_2}_{0,b\bar{b}}+\left(\frac{\als}{\pi}\right)\Delta^{C_2C_2}_{1,b\bar{b}}+\left(\frac{\als}{\pi}\right)^2\Delta^{C_2C_2}_{2,b\bar{b}}+\left(\frac{\als}{\pi}\right)^3\Delta^{C_2C_2}_{3,b\bar{b}}
+\left(\frac{\als}{\pi}\right)^4\Delta^{C_2C_2}_{4,b\bar{b}}+\mathcal{O}(\als^5)\right],\nonumber\\
\Gamma^{C_1C_2}_{H\rightarrow b\bar{b}} &= C_1C_2\left[\left(\frac{\als}{\pi}\right)\Delta^{C_1C_2}_{1,b\bar{b}}+\left(\frac{\als}{\pi}\right)^2\Delta^{C_1C_2}_{2,b\bar{b}}
+\left(\frac{\als}{\pi}\right)^3\Delta^{C_1C_2}_{3,b\bar{b}}+\mathcal{O}(\als^4)\right],\nonumber\\
\Gamma^{C_1C_1}_{H\rightarrow b\bar{b}} &= C_1C_1\left[\left(\frac{\als}{\pi}\right)\Delta^{C_1C_1}_{1,b\bar{b}}+
\left(\frac{\als}{\pi}\right)^2\Delta^{C_1C_1}_{2,b\bar{b}}
+\mathcal{O}(\als^3)\right],
\label{Delta}
\end{align}
where we have shown explicitly the terms needed to obtain $\Gamma_{H\rightarrow b\bar{b}}$ to $\mathcal{O}(\als^4)$. 
Notice that the leading order contributions of $\Gamma^{C_1C_2}_{H\rightarrow b\bar{b}}$ and $\Gamma^{C_1C_1}_{H\rightarrow b\bar{b}}$ are $\mathcal{O}(\als^2)$ and $\mathcal{O}(\als^3)$, respectively.

The results of $\Delta_{i,b\bar{b}}^{C_2 C_2}$ have been obtained up to $\mathcal{O}(\alpha_s^4)$ with massless final-state bottom quarks \cite{Gorishnii:1990zu,Chetyrkin:1996sr,Baikov:2005rw,Herzog:2017dtz}.
Defining $z\equiv m_H^2/m_b^2$,
they are given by
\begin{align}
\Delta^{C_2C_2}_{0,b\bar{b}}|_{z\to \infty} &= \frac{3m_H\overline{m_b}(m_H)^2}{8v^2\pi }+\mathcal{O}(z^{-1}),  \nonumber\\
\Delta^{C_2C_2}_{1,b\bar{b}}|_{z\to \infty} &= \frac{17m_H\overline{m_b}(m_H)^2}{8v^2\pi }+\mathcal{O}(z^{-1}),  \nonumber\\
\Delta^{C_2C_2}_{2,b\bar{b}}|_{z\to \infty} &= \frac{m_H\overline{m_b}(m_H)^2}{96v^2\pi }\bigg[
-582 \zeta (3)-47 \pi ^2+\frac{8851}{4}
\bigg]+\mathcal{O}(z^{-1}),  \nonumber\\
\Delta^{C_2C_2}_{3,b\bar{b}}|_{z\to \infty} &=\frac{m_H\overline{m_b}(m_H)^2}{96v^2\pi }  
\bigg[
1945 \zeta (5)-\frac{5 \pi ^4}{3}-\frac{80095 \zeta (3)}{6}-\frac{10225 \pi ^2}{9}+\frac{34873057}{1296}
\bigg] +\mathcal{O}(z^{-1}),  \nonumber\\
\Delta^{C_2C_2}_{4,b\bar{b}}|_{z\to \infty} &= 
\frac{m_H\overline{m_b}(m_H)^2}{96v^2\pi }  
\bigg[
-\frac{427735 \zeta (7)}{64}+\frac{50 \pi ^6}{189}+\frac{84625 \zeta (3)^2}{3}+\frac{469675 \zeta (5)}{12}
+\frac{116945\pi ^2 \zeta (3)}{24}
\nonumber\\&\quad
+\frac{667 \pi ^4}{8}-\frac{12308897 \zeta (3)}{48}-\frac{19637651 \pi ^2}{864}+\frac{5005075879}{13824}
\bigg]+\mathcal{O}(z^{-1}).
\label{eq:c2c2}
\end{align}
The analytical result with full bottom-quark mass dependence at $\mathcal{O}(\alpha_s^2)$ was obtained in \cite{Wang:2023xud}, 
showing that the $\mathcal{O}(z^{-1})$ power correction is below $1\%$ relative to the leading power result.
This is because  the expansion parameter  is rather small ($z^{-1}\approx 0.1\%$).
Similar corrections are expected at higher orders.
Since the leading power corrections at $\mathcal{O}(\alpha_s^3)$ and $\mathcal{O}(\alpha_s^4)$ are already small (at the per-mille level),
the $\mathcal{O}(z^{-1})$ power corrections at these orders are thus completely negligible.

\begin{figure}[ht]
	\centering
\begin{minipage}{0.3\linewidth}
	\centering
	\includegraphics[width=1\linewidth]{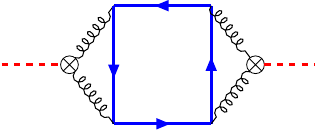}
 \caption*{(a)}
\end{minipage}
\,\,
 \begin{minipage}{0.3\linewidth}
		\centering
    \includegraphics[width=1\linewidth]{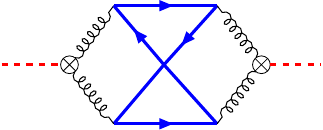}
   \caption*{(b)}
\end{minipage}
\caption{Sample three-loop Feynman diagrams contributing to $\Delta^{C_1C_1}_{2,b\bar{b}}$. The thick blue and dashed red lines denote the massive bottom quark and the Higgs boson, respectively.}
\label{ThreeLoop}
\end{figure}

The full analytic results of $\Delta_{i,b\bar{b}}^{C_1 C_2}$ with $i=1,2$ can be found in Ref. \cite{Wang:2024ilc}.
In the small $m_b$ limit, ${\Delta}^{C_1C_2}_{1,b\bar{b}}$ is logarithmically enhanced, given its asymptotic form
\begin{align}
{\Delta}^{C_1C_2}_{1,b\bar{b}}|_{z \rightarrow \infty} &= \frac{4m_Hm_b\overline{m_b}(\mu)}{\pi v^2}\bigg[-\frac{1}{8}\log^2(z)-\frac{3}{4}\log\left(\frac{\mu ^2}{m_H^2}\right)
+\frac{\pi^2}{8}-\frac{19}{8}
\nonumber\\&
+\frac{1}{2}\frac{\log ^2(z)}{z}+2\frac{\log (z)}{z}+\frac{9}{2 z} \log \left(\frac{\mu ^2}{m_H^2}\right)-\frac{\pi ^2}{2 z}+\frac{15}{2 z}\bigg]+\mathcal{O}(z^{-2}).
\label{eq:deltac1c2}
\end{align}
The prefactor $\overline{m_b}(\mu)$ comes from the bottom-quark Yukawa coupling,
while $m_b$ arises from the bottom-quark propagators. 
The expression of ${\Delta}^{C_1C_2}_{2,b\bar{b}}$ is  too lengthy to be shown here.
A prominent feature is that it contains logarithmic  terms up to $\log^4(z)$.
So far, a calculation of ${\Delta}^{C_1C_2}_{3,b\bar{b}}$, either analytical or numerical, is still missing.

The analytic result of $\Delta_{1,b\bar{b}}^{C_1 C_1}$ has also been known in \cite{Wang:2024ilc}, and the asymptotic form is 
\begin{align}{\Delta}^{C_1C_1}_{1,b\bar{b}}|_{z \rightarrow \infty} & = \frac{4m_H^3}{\pi v^2}
\bigg[\frac{1}{6}  \log (z)-\frac{7}{12}+\frac{3}{z}\bigg]
+\mathcal{O}(z^{-2})\,.
\label{eq:c1c1}
\end{align}
It can be seen that this contribution is power enhanced compared to the results of $\Delta_{i,b\bar{b}}^{C_2 C_2}$.
Due to these features shown in eqs. (\ref{eq:deltac1c2},\ref{eq:c1c1}),
the top-quark Yukawa coupling induced contributions cannot be obtained 
if the final-state bottom quarks are assumed to be massless.
In this work, we focus on the contribution of $\Gamma^{C_1C_1}_{H\rightarrow b\bar{b}}$ at $\mathcal{O}(\als^4)$,
  then we need to calculate $\Delta^{C_1C_1}_{2,b\bar{b}}$.
We perform the calculation using the optical theorem, i.e.,
\begin{align}
\Gamma_{H\rightarrow b\bar{b}}  = \frac{\text{Im}_{b\bar{b}}\left(\Sigma\right)}{m_H}  ,
\label{eq:optical}
\end{align}
where $\Sigma$ represents the amplitude of  $H \rightarrow b \bar{b} +X \rightarrow H$ with $X$ denoting any particles.
Specifically, $\Delta^{C_1C_1}_{2,b\bar{b}}$ corresponds to the three-loop Feynman diagrams which contain two $\mathcal{O}_1$ vertices.  Typical diagrams can be seen in figure \ref{ThreeLoop}.
The notation ${\rm Im}_{b\bar{b}}$ in eq. (\ref{eq:optical}) indicates that the imaginary part is taken only for the cut on at least one bottom quark pair. 
Explicitly, $\Delta^{C_1C_1}_{2,b\bar{b}}$ can be divided into two parts
\begin{align}
{\Delta}^{C_1C_1}_{2,b\bar{b}}   = \tilde{\Delta}^{C_1C_1}_{2,b\bar{b}}+ {\Delta}^{C_1C_1}_{2,b\bar{b}b\bar{b}},   
\label{eq:div}
\end{align}
where $\tilde{\Delta}^{C_1C_1}_{2,b\bar{b}}$ represents the contribution from the final states of $b\bar{b}$, $b\bar{b}g$, $b\bar{b}gg$, and $b\bar{b}q\bar{q}$,
while ${\Delta}^{C_1C_1}_{2,b\bar{b}b\bar{b}}$ receives contributions from the final state of $b\bar{b}b\bar{b}$.

The Feynman diagrams are generated using the package {\tt FeynArts} \cite{Hahn:2000kx} with effective vertices being implemented via {\tt FeynRules} \cite{Alloul:2013bka}.
The corresponding amplitudes are simplified 
with the package {\tt FeynCalc} \cite{Shtabovenko:2020gxv,Shtabovenko:2023idz} and expressed as a linear combination of scalar integrals.
They are reduced to a set of basis integrals called master integrals (MIs) using the integration by parts (IBP) identities \cite{Tkachov:1981wb,Chetyrkin:1981qh}  with the help of the package {\tt Kira} \cite{Klappert:2020nbg}.

\section{Analytic calculation}
\label{sec:MIs}

We find that the MIs belong to one integral family,
which is defined by
\begin{align}
	I_{n_1,n_2,\ldots,n_{9}}=\textrm{Im}_{b\bar{b}} \int \prod_{i=1}^3  \frac{(m_b^2)^{\eps}d^{4-2\eps} q_i}{i\pi^{2-\eps}\Gamma(1+\eps)}~\frac{1}{D_1^{n_1}~D_2^{n_2}~D_3^{n_3}~D_4^{n_4}~D_5^{n_5}~D_6^{n_6}~D_7^{n_7}~D_8^{n_8}~D_9^{n_9}}\,, 
\end{align}
with all $n_i$ being integers. 
The denominators $D_i$ are defined as
\begin{align}
	D_1 &= q_1^2,&
	D_2 &= (q_1-q_2)^2-m_b^2,&
	D_3 &= q_2^2-m_b^2,\nonumber\\
	D_4 &= q_3^2,&
	D_5 &= (q_2+q_3)^2-m_b^2,&
	D_6 &= (q_3-k)^2-m_b^2,\nonumber\\
	D_7 &= (q_1+k)^2,&
	D_8 &=(q_1-q_2-q_3+k)^2,&
	D_9 &=(q_1-q_2-q_3)^2-m_b^2,
\end{align}
where the momentum of the Higgs boson satisfies $k^2 = m_H^2$.
There are 38 MIs ($\M_1$-$\M_{38}$) appearing in the calculations of amplitudes.
The corresponding topological diagrams are shown in appendix \ref{sec:topo}.
 Similar to eq. (\ref{eq:div}), the MIs can be divided into two parts,
\begin{align}
\M_{i} =  {\tilde\M}_{i,b\bar{b}} + \M_{i,4b}, 
\end{align}
where ${\tilde\M}_{i,b\bar{b}}$ and  $\M_{i,4b}$ contribute to  $\tilde{\Delta}^{C_1C_1}_{2,b\bar{b}}$ and ${\Delta}^{C_1C_1}_{2,b\bar{b}b\bar{b}}$, respectively. 

We find that $\M_{1}$-$\M_{35}$ have been analytically calculated in our previous works \cite{Chen:2024amk,Wang:2023xud,Wang:2024ilc}. 
Most of these MIs can contribute to $\tilde{\Delta}^{C_1C_1}_{2,b\bar{b}}$ except for the first three MIs (banana diagrams at three loops), which only contain cuts on $b\bar{b}b\bar{b}$, i.e.,
\begin{align}
{\tilde\M}_{1,b\bar{b}} = {\tilde\M}_{2,b\bar{b}} = {\tilde\M}_{3,b\bar{b}} = 0.   
\end{align}
Since ${\M}_{1,4b}$-${\M}_{3,4b}$, containing elliptic integrals, serve as the lower sector of the other $\M_{i,4b}$, all $\M_{i,4b}$ are consequently related to the elliptic integrals. Fortunately, $\Delta^{C_1C_1}_{2,b\bar{b}b\bar{b}}$ is finite as it corresponds to the interference of tree-level diagrams for $H\rightarrow b\bar{b}b\bar{b}$. By adopting a regular basis in \cite{Lee:2019wwn}, only the $\mathcal{O}(\eps^0)$ parts are required in the amplitude calculations.  Finally, all $\M_{i,4b}~(i=1,\cdots,35)$ can be expressed either as complete elliptic integrals of the first kind or as one-fold integrals of them.   
The integrals ${\tilde\M}_{i,b\bar{b}}~(i=1,\cdots,35)$ can be solved by constructing canonical differential equations \cite{Kotikov:1991pm,Henn:2013pwa}. 
In this procedure, there are two square roots
\begin{align}
r_1 = \sqrt{z(z-4)}, \quad    r_2 = \sqrt{z(z+4)}.
\end{align}
The variable redefinition 
\begin{align}
z = -\frac{(w-1)^2}{w}, \quad -1<w<0    
\end{align}
can rationalize $r_1$ while 
\begin{align}
z = -\frac{\left(y^2+1\right)^2}{(y-1) y (y+1)}, \quad 0<y<\sqrt{2}-1
\end{align}
can rationalize both $r_1$ and $r_2$. 
In this way, all ${\tilde\M}_{i,b\bar{b}}~(i=1,\cdots,35)$ can be solved recursively, yielding analytic results expressed as linear combinations of multiple polylogarithms (MPLs) \cite{Goncharov:1998kja}. 
The boundary conditions are fixed by the PSLQ algorithm \cite{Ferguson:1999aa} together with the high-precision numerical results obtained from the package {\tt AMFlow} \cite{Liu:2017jxz,Liu:2020kpc,Liu:2022chg,Liu:2022mfb}. 

$\M_{36}$-$\M_{38}$ are the new MIs to be computed,
\begin{align}
\M_{36} =I_{1,1,0,1,1,1,1,1,1}\,,\quad
\M_{37} =I_{1,1,1,1,1,1,1,1,0}\,,\quad
\M_{38} =I_{1,1,1,1,1,1,1,2,0}\,.    
\end{align}
$\M_{36}$ only contributes to $\tilde{\Delta}^{C_1C_1}_{2,b\bar{b}}$, i.e., $\M_{36,4b}$ = 0. We find that 
$(1-2\epsilon)\epsilon^5m_b^4r_1z\M_{36} $
forms a dimensionless canonical basis, which can be expressed in terms of MPLs using the variable $y$.

We now turn to the top sector consisting of $\M_{37}$ and $\M_{38}$. 
These two MIs contribute to both $\tilde{\Delta}^{C_1C_1}_{2,b\bar{b}}$ and ${\Delta}^{C_1C_1}_{2,b\bar{b}b\bar{b}}$. In both cases, they are finite, and thus only their $\mathcal{O}(\epsilon^0)$ parts are required. Accordingly, we define
\begin{align}
F_{37,b\bar{b}} &= z^2 m_b^4\tilde{\M}_{37,b\bar{b}}|_{\eps=0}\,,\quad
F_{38,b\bar{b}} = z^2 m_b^6\tilde{\M}_{38,b\bar{b}}|_{\eps=0}\,,\nonumber\\
F_{37,4b} &= z^2 m_b^4\M_{37,4b}|_{\eps=0}\,,\quad
F_{38,4b} = z^2 m_b^6\M_{38,4b}|_{\eps=0}\,.
\end{align}
These basis integrals satisfy the differential equations
\begin{align}
\frac{\partial F_{37,b\bar{b}\,(4b)}}{\partial z} &= -\frac{4  F_{38,b\bar{b}\,(4b)}}{z},\nonumber\\
\frac{\partial F_{38,b\bar{b}\,(4b)}}{\partial z} &= \frac{F_{37,b\bar{b}\,(4b)}}{z (z+16)}+\frac{16 F_{38,b\bar{b}\,(4b)} }{z (z+16)}+R_{b\bar{b}\,(4b)}(z),
\end{align}
where the homogeneous parts are the same for the $b\bar{b}$ and $4b$ cuts.
The only difference is represented by $R_{b\bar{b}(4b)}(z)$,
 which is fully determined by the MIs in lower sectors. Their solutions can be written as one-fold integrals:
\begin{align}
F_{37,b\bar{b}}(z) &= \int^z_{4}
\frac{\left(K\left(-\frac{x}{16}\right) K\left(\frac{z+16}{16}\right)-K\left(\frac{x+16}{16}\right) K\left(-\frac{z}{16}\right)\right)(x+16)\sqrt{ z}}{\pi \sqrt{x}}R_{b\bar{b}}(x)dx,\\
F_{38,b\bar{b}}(z) &= \int^z_{4}
\frac{2\left(K\left(\frac{x+16}{16}\right) E\left(-\frac{z}{16}\right)+K\left(-\frac{x}{16}\right) \left(E\left(\frac{z+16}{16}\right)-K\left(\frac{z+16}{16}\right)\right)\right)(x+16)\sqrt{z}}{\pi(z+16)\sqrt{x}}R_{b\bar{b}}(x)dx,  \nonumber
\end{align}
and similarly
\begin{align}
F_{37,4b}(z) &= \int^z_{16}
\frac{\left(K\left(-\frac{x}{16}\right) K\left(\frac{z+16}{16}\right)-K\left(\frac{x+16}{16}\right) K\left(-\frac{z}{16}\right)\right)(x+16)\sqrt{ z}}{\pi\sqrt{x}}R_{4b}(x)dx,\\
F_{38,4b}(z) &= \int^z_{16}
\frac{2\left(K\left(\frac{x+16}{16}\right) E\left(-\frac{z}{16}\right)+K\left(-\frac{x}{16}\right) \left(E\left(\frac{z+16}{16}\right)-K\left(\frac{z+16}{16}\right)\right)\right)(x+16)\sqrt{z}}{\pi(z+16)\sqrt{x}}R_{4b}(x)dx,  \nonumber
\end{align}
where $K$ and $E$ are the complete elliptic integrals of the first and second kinds, respectively. The distinctions between $F_{i,b\bar{b}}(z)$ and $F_{i,4b}(z)$ lie in the lower integration limits and the specific forms of $R(x)$. It can be seen that these integrals vanish at the threshold, i.e., $z=4(16)$ for $F_{i,b\bar{b}}(z)$ ($F_{i,4b}(z)$). For the $b\bar{b}$ final state, $R_{b\bar{b}}$ is the linear combination of logarithmic and $\text{Li}_2$ functions, so the final results of $F_{i,b\bar{b}}(z)$ are expressed as one-fold integrals. For the $b\bar{b}b\bar{b}$ final state, however, $R_{4b}$ itself is already a one-fold integral \cite{Lee:2019wwn}. Consequently, the final expressions of $F_{i,4b}(z)$ take the form of two-fold integrals. 

The above integral representation can be numerically evaluated. 
In addition, we can derive the asymptotic form in the small $m_b$ ($z\rightarrow\infty$) limit following the method in \cite{Lee:2020obg}, 
\begin{align}
F_{37,b\bar{b}}\,|_{z \rightarrow \infty} &=  -\frac{11\pi}{6}\log ^4(z)+\frac{31\pi ^3}{3}\log ^2(z)-128 \pi  \zeta (3) \log (z)-\frac{323 \pi ^5}{90}+\mathcal{O}(z^{-1}),\nonumber\\
F_{38,b\bar{b}}\,|_{z \rightarrow \infty} &=\frac{11 \pi}{6}\log ^3(z)-\frac{31 \pi^3 }{6}\log(z)+32 \pi  \zeta (3)+\mathcal{O}(z^{-1}),\nonumber\\
F_{37,4b}\,|_{z \rightarrow \infty} &=
\frac{7\pi}{6}\log ^4(z)-\frac{13\pi ^3}{3}\log ^2(z)+48\pi\zeta(3)\log(z)+\frac{181 \pi ^5}{90}+\mathcal{O}(z^{-1}),\nonumber\\
F_{38,4b}\,|_{z \rightarrow \infty} &=-\frac{7 \pi}{6}\log ^3(z)+\frac{13\pi^3 }{6}\log(z)-12 \pi  \zeta (3)+\mathcal{O}(z^{-1}).
\end{align}

Before proceeding, we comment on another method to compute the MIs.
We have constructed the $\epsilon$-factorised differential equations for all master integrals, including the banana integrals and the above top sector,  which is elliptic by itself, and solved them beyond leading orders in $\eps$. 
We find full agreement between these two methods.
The details will be presented in a forthcoming paper \cite{Wang:2026ans},
and the upshot is as follows. Firstly, the solutions in expansions of $\eps$ are straightforward to obtain and do not involve one-fold or two-fold integrals. Secondly, this integral family, and hence the physical observable, incorporates several geometric objects in a non-trivial way. Consequently, it is a  practical application of the algorithm to find the $\eps$-factorised differential equation proposed recently in \cite{e-collaboration:2025frv,Bree:2025tug}, including mixing between different sectors. On top of that, since all the integrals involve only one dimensionless variable $z$, it is potentially a good playground to investigate how different geometries talk to each other at the observable level.

All the MIs have been numerically cross checked with the package {\tt AMFlow} \cite{Liu:2017jxz,Liu:2020kpc,Liu:2022chg,Liu:2022mfb} at several fixed points, such as $z = 625$.
After computing the MIs, we obtain the analytical results for the three-loop cut amplitude, which still contain ultraviolet divergences.
One has to  perform renormalization to obtain finite results.
This requires considering the contribution from renormalization constants $Z_{ij}$ in eq. (\ref{eq:renZ}) in addition to the usual mass and coupling renormalization.
See refs. \cite{Wang:2023xud,Wang:2024ilc} for more details.
The full results of the decay width are provided in the auxiliary file.
Here, we only present the asymptotic form:
\begin{align}
&\tilde{\Delta}^{C_1C_1}_{2,b\bar{b}}|_{z \rightarrow \infty} = \frac{m_H^3}{\pi v^2}\times\nonumber\\
&\bigg(C_AC_F^2\Big[\frac{\log (z)}{8}+\frac{\zeta (3)}{3}-\frac{29}{96}-\frac{\log ^4(z)}{96 z}+\frac{\log ^3(z)}{8 z}+\frac{3}{8}\log \left(\frac{\mu ^2}{m_H^2}\right)\frac{\log ^2(z)}{z}
\nonumber\\&
-\frac{\pi ^2}{48}\frac{\log^2(z)}{z}+\frac{3}{8}\frac{\log ^2(z)}{z}+\frac{5 \zeta (3)}{2}\frac{\log (z)}{z}+\frac{13 \pi ^2}{24}\frac{\log (z)}{z}-\frac{41}{8}\frac{\log (z)}{z}
\nonumber\\&
+\frac{9}{8 z}\log ^2\left(\frac{\mu ^2}{m_H^2}\right)
-\frac{3 \pi ^2}{8 z}\log \left(\frac{\mu ^2}{m_H^2}\right)+\frac{57}{8 z}\log \left(\frac{\mu ^2}{m_H^2}\right)+\frac{3 \pi ^4}{160 z}-\frac{15 \zeta (3)}{2 z}-\frac{7 \pi ^2}{12 z}+\frac{237}{16 z}\Big]
\nonumber\\&
+C_A^2C_F\Big[
\frac{\log^3(z)}{72}-\frac{5\log^2(z)}{72}-\frac{\pi ^2}{18}\log(z)+\frac{11}{24}\log \left(\frac{\mu^2}{m_H^2}\right)\log(z)+\frac{859 \log (z)}{432}
\nonumber\\&
-\frac{77}{48}\log \left(\frac{\mu ^2}{m_H^2}\right)
+\frac{\zeta (3)}{4}+\frac{53 \pi ^2}{216}-\frac{4897}{648}+\frac{\log ^3(z)}{12 z}+\frac{3 \log ^2(z)}{4 z}-\frac{7 \pi ^2}{24}\frac{\log (z)}{z}
\nonumber\\&
+\frac{\log (z)}{8 z}+\frac{33}{4 z}\log \left(\frac{\mu ^2}{m_H^2}\right)+\frac{4 \zeta (3)}{z}-\frac{11 \pi ^2}{12 z}+\frac{457}{16 z}
\Big]
\nonumber\\&
+C_AC_Fn_l\Big[
-\frac{\log ^2(z)}{36}-\frac{7 \log (z)}{72}-\frac{1}{12}\log \left(\frac{\mu ^2}{m_H^2}\right)\log(z)+\frac{7}{24}\log \left(\frac{\mu ^2}{m_H^2}\right)
\nonumber\\&
+\frac{455}{648}-\frac{\log (z)}{z}-\frac{3}{2 z}\log \left(\frac{\mu ^2}{m_H^2}\right)-\frac{9}{4 z}
\Big]
\nonumber\\&
+C_AC_F\Big[
-\frac{\log ^2(z)}{18}+\frac{7 \log (z)}{72}-\frac{1}{12}\log \left(\frac{\mu ^2}{m_H^2}\right)\log(z)+\frac{7}{24}\log \left(\frac{\mu ^2}{m_H^2}\right)-\frac{\pi^2}{108}
\nonumber\\&
+\frac{245}{648}-\frac{\log (z)}{2z}-\frac{3}{2 z}\log \left(\frac{\mu ^2}{m_H^2}\right)-\frac{13}{6 z}
\Big]\bigg)+\mathcal{O}(z^{-2}),
\end{align}
and
\begin{align}
&\Delta^{C_1C_1}_{2,b\bar{b}b\bar{b}}|_{z \rightarrow \infty} = \frac{m_H^3}{\pi v^2}\times\nonumber\\
&\bigg(\left(2C_AC_F^2-C_A^2C_F\right)\Big[
\frac{5}{96}-\frac{\zeta (3)}{24}+\frac{\log ^2(z)}{16 z}-\frac{3 \log (z)}{16 z}-\frac{\pi ^2}{16 z}+\frac{7}{32 z}\Big]
\nonumber\\&
+C_AC_F\Big[\frac{\log ^2(z)}{72}-\frac{7 \log (z)}{72}-\frac{\pi ^2}{216}+\frac{17}{81}+\frac{\log (z)}{2 z}-\frac{19}{12 z}\Big]\bigg)+\mathcal{O}(z^{-2})\,.
\end{align}
In fact, these asymptotic results are accurate enough for phenomenological studies since the omitted higher-power terms introduce  a correction of less than 0.1\%. 

One would wonder whether it is possible to derive these asymptotic results directly by expanding the amplitude in eq. (\ref{eq:optical}).
The resulting integrals are usually evaluated with the method of regions \cite{Smirnov:1997gx,Beneke:1997zp}.
However, its direct application to loop integrals with cuts on propagators is not yet widely established.
One can also resort to the expansions of the differential equations near singular points \cite{Lee:2017qql}.
Although the logarithmic terms can be fully determined from the differential equations, the constant terms in each coefficient of $z^{-n}$ still have to be calculated by another method. 

The above results are given in terms of the on-shell mass $m_b$.
They can be converted to the $\overline{\text{MS}}$ scheme  via the relation 
\begin{align}
\overline{m_b}(\mu) = m_b\left( 1 - \left(\frac{\als}{\pi}\right)C_F\left[1+\frac{3}{4}\log \left(\frac{\mu ^2}{m_b^2}\right)\right]+\mathcal{O}(\als^2)\right)\,.
\end{align}
Compared to the contributions induced by the bottom-quark Yukawa coupling in eq. (\ref{eq:c2c2}), 
$\tilde{\Delta}^{C_1C_1}_{2,b\bar{b}}$ and $\Delta^{C_1C_1}_{2,b\bar{b}b\bar{b}}$ are power enhanced.
This feature has been previously observed in the calculation of ${\Delta}^{C_1C_1}_{1,b\bar{b}}$ \cite{Wang:2024ilc}; see eq. (\ref{eq:c1c1}).
The ratio of the leading logarithms of ${\Delta}^{C_1C_1}_{2,b\bar{b}}$ over that of ${\Delta}^{C_1C_1}_{1,b\bar{b}}$ is
\begin{align}
\frac{1}{12}C_A\log^2 (z)
\end{align}
at the leading power
and
\begin{align}
-\frac{1}{288}C_F\log^4 (z)
\end{align}
at the subleading power.
The distinct color structures and powers indicate that they have different origins.
The double-logarithmic enhancement arises from the diagram with a gluon splitting into a bottom-quark pair,
while the quartic term appears in the diagram featuring exchanges of two soft bottom quarks.
Similar enhancements have been found in the calculation of $\mathcal{O}(\alpha_s^2)$ corrections in the $C_2C_2$ channel with $b\bar{b}$ cuts \cite{Wang:2023xud} (see figure 8 there). 
The behaviour of such large logarithms at all orders is very interesting and deserves a detailed study in the future.

\section{Numerical results}
\label{sec:num}
To present the numerical results for the decay width of $H\rightarrow b\bar{b}$, we adopt the following input parameters \cite{ParticleDataGroup:2024cfk,ATLAS:2015yey}:
\begin{align}
    \overline{m_b}(\overline{m_b}) &= 4.18~\mathrm{GeV},   & m_H &= 125.09~\mathrm{GeV}, &m_t &= 172.57~\mathrm{GeV}, \nonumber\\
     \alpha_s(m_Z) &= 0.1180\pm 0.0009, & G_F &= 1.166378 \times 10^{-5}~\mathrm{GeV}^{-2}.
    \label{eq:input}
\end{align} 
$\overline{m_b}$ at other scales is evaluated with the package {\tt RunDec} \cite{Chetyrkin:2000yt,Herren:2017osy}, 
e.g., 
$\overline{m_b}\,(m_H/2) = 2.959$ GeV, $\overline{m_b}\,(m_H) = 2.787$ GeV and $\overline{m_b}\,(2m_H) = 2.642$ GeV. 
The running strong coupling at the typical scales reads $\als\,(m_H/2) = 0.1251$, $\als\,(m_H) = 0.1126$ and $\als\,(2m_H) = 0.1024$.

In table \ref{tab:width}, we show the different contributions to the decay width $\Gamma_{H\to b\bar{b}}$ in the $\overline{\text{MS}}$ scheme.
The LO and NLO results come only from the $C_2C_2$ channel.
At NNLO, the $C_1C_2$ channel provides a correction of $0.8\%$, which is one-fourth of that in the $C_2C_2$ channel.
At NNNLO, the sum of the $C_1C_2$ and $C_1 C_1$ channels is about five times that of the $C_2C_2$ channel, signifying the importance of including the top-quark Yukawa contributions.
The $\mathcal{O}(\alpha_s^4)$ corrections in the $C_2C_2$ channel decrease the decay width by $0.1\%$.
In contrast, the $\mathcal{O}(\alpha_s^4)$ corrections in the $C_1C_1$ channel still enhance the decay width by $0.4\%$,
a magnitude larger than the expected experimental precision ($0.21\%$) at future lepton colliders \cite{Altmann:2025feg}.

\begin{table}[H]
	\centering
	\scalebox{1.0}{
		\begin{tabular}{clccc}
			\toprule
            $\mu=m_H$&
			[MeV]&
			$\Gamma^{C_2C_2}_{Hb\bar{b}}$\quad &
			$\Gamma^{C_1C_2}_{Hb\bar{b}}$\quad &
			$\Gamma^{C_1C_1}_{Hb\bar{b}}$ \quad\\
			\midrule
            \multirow{6}{*}{$\Gamma_{H\rightarrow b\bar{b}}\left(\overline{\rm MS}\right)$}&
			$\mathcal{O}(\als^0)$&
            1.9076&
			-&
			-\\
			\cmidrule{2-5}
   		&$\mathcal{O}(\als^1)$&
            0.3873&
			-&
			-\\
			\cmidrule{2-5}
      	&$\mathcal{O}(\als^2)$&
            0.0735&
			0.0183&
			-\\
			\cmidrule{2-5}
            &$\mathcal{O}(\als^3)$&
            0.0048&
			0.0142&
			0.0090\\
   		\cmidrule{2-5}
            &$\mathcal{O}(\als^4)$&
            -0.0025&
			$*$&
			0.0087\\
			\midrule   
		\end{tabular}}
	\caption{Different contributions to the decay width in the $\overline{\rm MS}$ scheme at $\mu=m_H$.  The notation `$*$' represents a
correction which has not been calculated.} 
	\label{tab:width}
\end{table}

\begin{figure}[ht]
	\centering
	\includegraphics[width=0.6\linewidth]{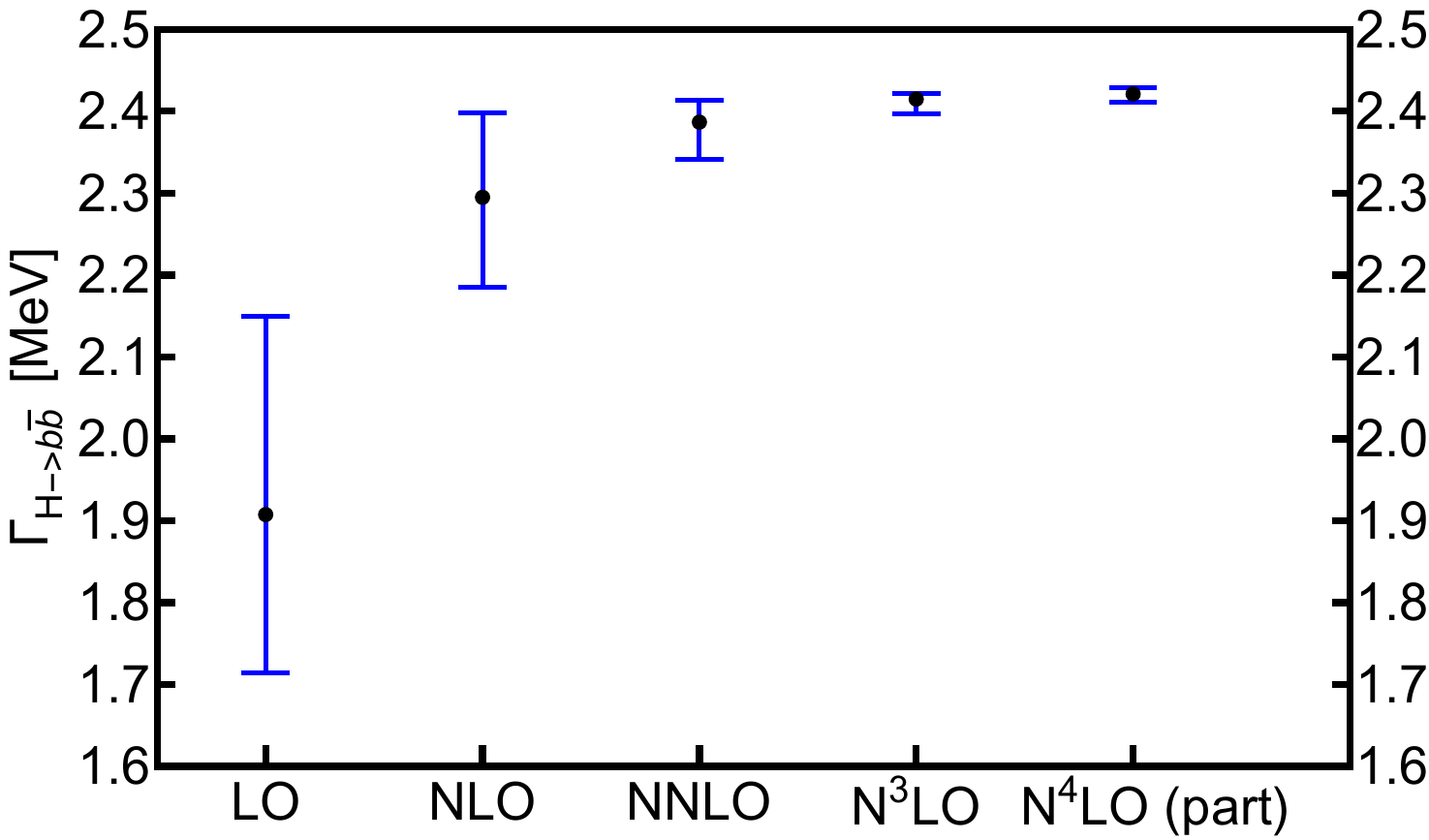}
\caption{The decay width of $H\to b\bar{b}$ in the $\overline{\rm MS}$ scheme at different perturbative orders. The error bar denotes the scale uncertainty.}
\label{fig:scale}
\end{figure}

\begin{figure}[ht]
	\centering
	\includegraphics[width=0.6\linewidth]{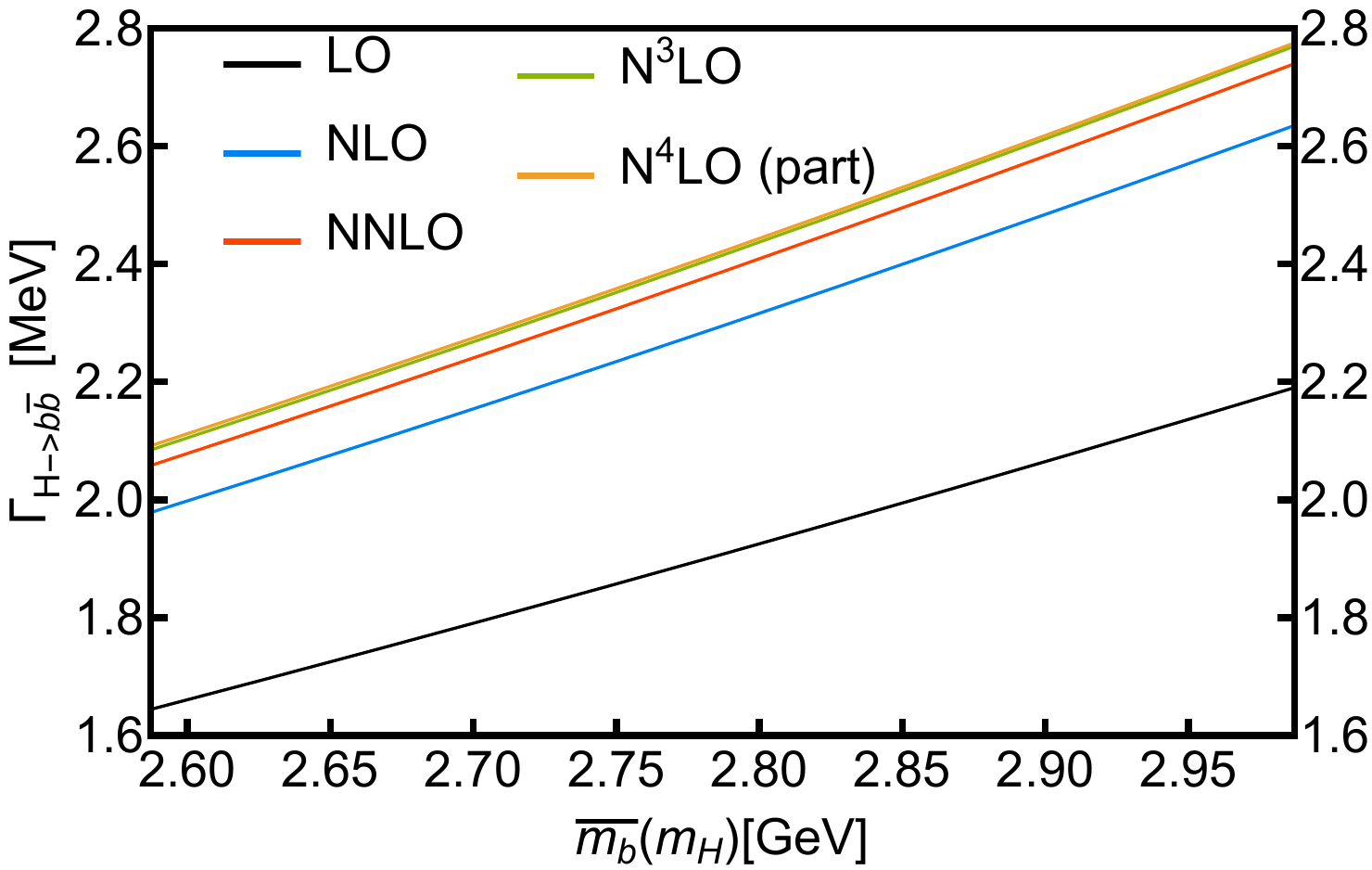}
\caption{The decay width of $H\to b\bar{b}$  as a function of the bottom quark mass $\overline{m_b}\,(m_H)$ that varies from 2.584 GeV  to  2.984 GeV.}
\label{fig:diff_mb}
\end{figure}

The scale uncertainty of the decay width is estimated by varying the renormalization scale around its default value $m_H$ by a factor of two.
As shown in figure \ref{fig:scale}, the scale uncertainty is significantly improved, reducing from $0.7\%$ at NNNLO  to $0.4\%$ after including the $\mathcal{O}(\alpha_s^4)$ corrections we computed in this work.
In addition, the uncertainty from $\delta \alpha_s = 0.0009$ brings an error of $0.2\%$.
Therefore, we obtain 
\begin{align}
    \Gamma_{H\rightarrow b\bar{b}}^{\rm N^4LO ~QCD}\left(\overline{\rm MS}\right) = 2.421 ^{+0.008}_{-0.010}({\rm scl.})^{+0.005}_{-0.005}({\alpha_s}) {~\rm MeV}.
\end{align}

The decay width can be used to derive the value of the bottom quark mass.
In figure \ref{fig:diff_mb}, we show the relation between these two quantities.
If the decay width is measured with an uncertainty of $0.21\%$ \cite{Altmann:2025feg},
the bottom quark mass can be derived with a precision around $0.36\%$,
where the theoretical uncertainties are taken into account.

\section{Conclusion}
\label{sec:con}
The decay width of the Higgs boson to bottom quarks is a fundamental quantity that will be precisely measured at future colliders.
To match the experimental accuracy, it is essential to provide theoretical predictions with the same or even better precision.
We calculate the $\mathcal{O}(\alpha_s^4)$ corrections in the $C_1C_1$ channel, finding that
they increase the decay width by $0.4\%$
 and reduce the scale dependence to $0.4\%$.
We also discuss the uncertainty from the strong coupling and present the expected precision of the bottom quark mass derived from the measurements of the decay width at future lepton colliders.
The next goal would be the $\mathcal{O}(\alpha_s^4)$ corrections in the $C_1C_2$ channel in order to provide a theoretical prediction with sufficient accuracy.

\section*{Acknowledgements}
We thank Dong-Hao Li and Xiaofeng Xu for helpful discussions. 
This work was supported by the National Natural Science Foundation of China under Nos. 12405117, 12321005, 12375076, 12535006. X.W. is also supported by the University Development Fund of The Chinese University of Hong Kong, Shenzhen, under the Grant No. UDF01003912.

\appendix
\section{Topological diagrams of the master integrals}
\label{sec:topo}
\begin{figure}[H]
	\centering
	\begin{minipage}{0.15\linewidth}
		\centering
		\includegraphics[width=1\linewidth]{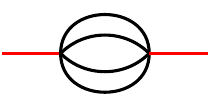}
		\caption*{$\M_{1}$}
	\end{minipage}
	\begin{minipage}{0.15\linewidth}
		\centering
		\includegraphics[width=1\linewidth]{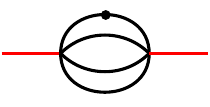}
		\caption*{$\M_{2}$}
	\end{minipage}
	\begin{minipage}{0.15\linewidth}
		\centering
		\includegraphics[width=1\linewidth]{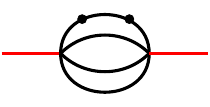}
		\caption*{$\M_{3}$}
	\end{minipage}
	\centering
	\begin{minipage}{0.15\linewidth}
		\centering
		\includegraphics[width=1\linewidth]{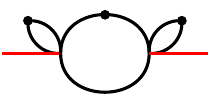}
		\caption*{$\M_{4}$}
	\end{minipage}
	\begin{minipage}{0.15\linewidth}
		\centering
		\includegraphics[width=1\linewidth]{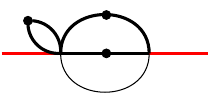}
		\caption*{$\M_{5}$}
	\end{minipage}
	\begin{minipage}{0.15\linewidth}
		\centering
		\includegraphics[width=1\linewidth]{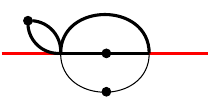}
		\caption*{$\M_{6}$}
	\end{minipage}
	\begin{minipage}{0.15\linewidth}
		\centering
		\includegraphics[width=1\linewidth]{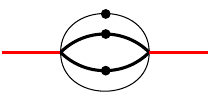}
		\caption*{$\M_{7}$}
	\end{minipage}
	\begin{minipage}{0.15\linewidth}
		\centering
		\includegraphics[width=1\linewidth]{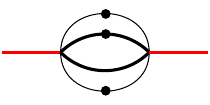}
		\caption*{$\M_{8}$}
	\end{minipage}
	\begin{minipage}{0.15\linewidth}
		\centering
		\includegraphics[width=1\linewidth]{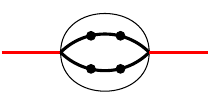}
		\caption*{$\M_{9}$}
	\end{minipage}
	\begin{minipage}{0.15\linewidth}
		\centering
		\includegraphics[width=1.1\linewidth]{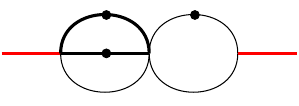}
		\caption*{$\M_{10}$}
	\end{minipage}
	\begin{minipage}{0.15\linewidth}
		\centering
		\includegraphics[width=1.1\linewidth]{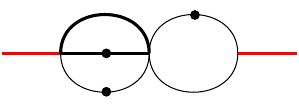}
		\caption*{$\M_{11}$}
	\end{minipage}
	\begin{minipage}{0.15\linewidth}
		\centering
		\includegraphics[width=1.1\linewidth]{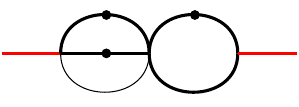}
		\caption*{$\M_{12}$}
	\end{minipage}
	\begin{minipage}{0.15\linewidth}
		\centering
		\includegraphics[width=1\linewidth]{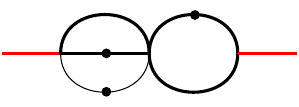}
		\caption*{$\M_{13}$}
	\end{minipage}
	\begin{minipage}{0.15\linewidth}
		\centering
		\includegraphics[width=1.0\linewidth]{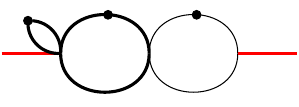}
		\caption*{$\M_{14}$}
	\end{minipage}
	\begin{minipage}{0.15\linewidth}
		\centering
		\includegraphics[width=1.0\linewidth]{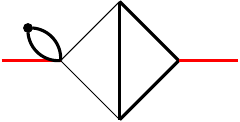}
		\caption*{$\M_{15}$}
	\end{minipage}
	\begin{minipage}{0.15\linewidth}
		\centering
		\includegraphics[width=1.0\linewidth]{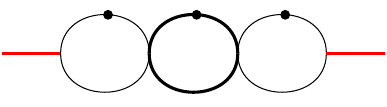}
		\caption*{$\M_{16}$}
	\end{minipage}
	\begin{minipage}{0.15\linewidth}
		\centering
		\includegraphics[width=1.0\linewidth]{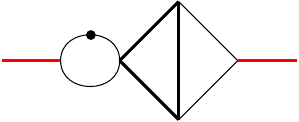}
		\caption*{$\M_{17}$}
	\end{minipage}
	\begin{minipage}{0.15\linewidth}
		\centering
		\includegraphics[width=1\linewidth]{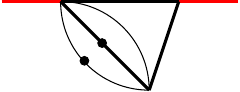}
		\caption*{$\M_{18}$}
	\end{minipage}
	\begin{minipage}{0.15\linewidth}
		\centering
		\includegraphics[width=1\linewidth]{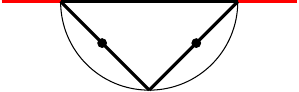}
		\caption*{$\M_{19}$}
	\end{minipage}
	\begin{minipage}{0.15\linewidth}
		\centering
		\includegraphics[width=1\linewidth]{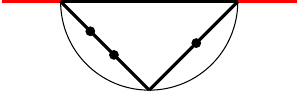}
		\caption*{$\M_{20}$}
	\end{minipage}
	\begin{minipage}{0.15\linewidth}
		\centering
		\includegraphics[width=1\linewidth]{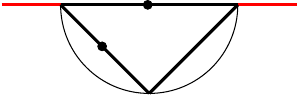}
		\caption*{$\M_{21}$}
	\end{minipage}
	\begin{minipage}{0.15\linewidth}
		\centering
		\includegraphics[width=1\linewidth]{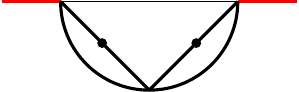}
		\caption*{$\M_{22}$}
	\end{minipage}
	\begin{minipage}{0.15\linewidth}
		\centering
		\includegraphics[width=1\linewidth]{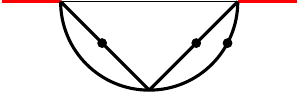}
		\caption*{$\M_{23}$}
	\end{minipage}
	\begin{minipage}{0.15\linewidth}
		\centering
		\includegraphics[width=1\linewidth]{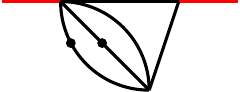}
		\caption*{$\M_{24}$}
	\end{minipage}
	\begin{minipage}{0.15\linewidth}
		\centering
		\includegraphics[width=1\linewidth]{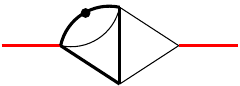}
		\caption*{$\M_{25}$}
	\end{minipage}
	\begin{minipage}{0.15\linewidth}
		\centering
		\includegraphics[width=1\linewidth]{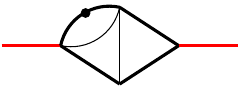}
		\caption*{$\M_{26}$}
	\end{minipage}
	\begin{minipage}{0.15\linewidth}
		\centering
		\includegraphics[width=1\linewidth]{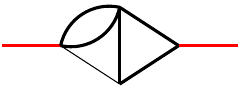}
		\caption*{$\M_{27}$}
	\end{minipage}
	\begin{minipage}{0.15\linewidth}
		\centering
		\includegraphics[width=1\linewidth]{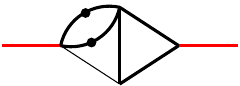}
		\caption*{$\M_{28}$}
	\end{minipage}
 	\begin{minipage}{0.15\linewidth}
		\centering
		\includegraphics[width=1\linewidth]{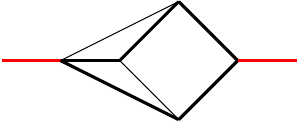}
		\caption*{$\M_{29}$}
	\end{minipage}
 	\begin{minipage}{0.15\linewidth}
		\centering
		\includegraphics[width=1\linewidth]{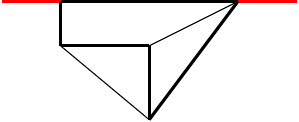}
		\caption*{$\M_{30}$}
	\end{minipage}
  	\begin{minipage}{0.15\linewidth}
		\centering
		\includegraphics[width=1\linewidth]{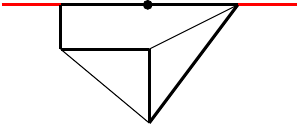}
		\caption*{$\M_{31}$}
	\end{minipage}
   	\begin{minipage}{0.15\linewidth}
		\centering
		\includegraphics[width=1\linewidth]{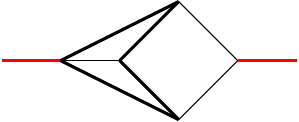}
		\caption*{$\M_{32}$}
	\end{minipage}
   	\begin{minipage}{0.15\linewidth}
		\centering
		\includegraphics[width=1\linewidth]{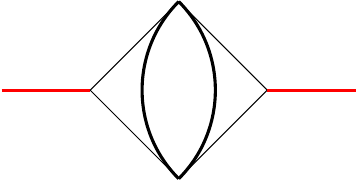}
		\caption*{$\M_{33}$}
	\end{minipage}
   	\begin{minipage}{0.15\linewidth}
		\centering
		\includegraphics[width=1\linewidth]{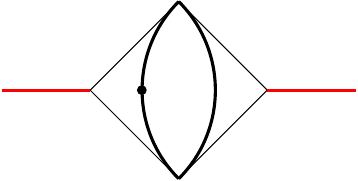}
		\caption*{$\M_{34}$}
	\end{minipage}
    	\begin{minipage}{0.15\linewidth}
		\centering
		\includegraphics[width=1\linewidth]{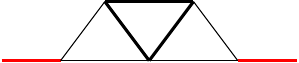}
		\caption*{$\M_{35}$}
	\end{minipage}
\begin{minipage}{0.15\linewidth}
		\centering
		\includegraphics[width=1\linewidth]{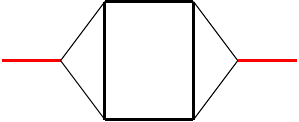}
		\caption*{$\M_{36}$}
\end{minipage}
\begin{minipage}{0.15\linewidth}
		\centering
		\includegraphics[width=1\linewidth]{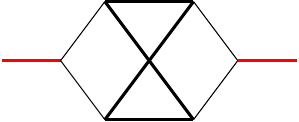}
		\caption*{$\M_{37}$}
\end{minipage}
\begin{minipage}{0.15\linewidth}
		\centering
		\includegraphics[width=1\linewidth]{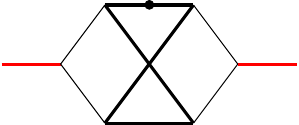}
		\caption*{$\M_{38}$}
\end{minipage}
\caption{ Topological diagrams of the master integrals. The thick black and red lines stand for the massive bottom quark and the Higgs boson, respectively. One black dot indicates one additional power of the corresponding propagator.}
\label{NP1_Topo}
\end{figure}

In this appendix, we show the topological diagrams of the master integrals in figure \ref{NP1_Topo}.

\bibliographystyle{JHEP}
\bibliography{reference}
\end{document}